\documentclass[twocolumn,showpacs,preprintnumbers,amsmath,amssymb]{revtex4}

\usepackage{graphicx}
\usepackage{dcolumn}
\usepackage{bm}

\newcommand{\bq}{\begin{equation}}
\newcommand{\eq}{\end{equation}}
\newcommand{\bqa}{\begin{eqnarray}}
\newcommand{\eqa}{\end{eqnarray}}
\newcommand{\nn}{\nonumber \\}

\def\be     {\begin{equation}}
\def\ee     {\end{equation}}
\def\bea        {\begin{eqnarray}}
\def\eea        {\end{eqnarray}}
\def\bnn    {\begin{eqnarray*}}
\def\enn    {\end{eqnarray*}}

\begin{document}

\title{Gr\"uneisen ratio at the Kondo breakdown quantum critical point}
\author{K.-S. Kim, A. Benlagra, and C. P\'epin}
\affiliation{Institut de Physique Th\'eorique, CEA, IPhT, CNRS,
URA 2306, F-91191 Gif-sur-Yvette, France}
\date{\today}

\begin{abstract}
We show that the scenario of multi-scale Kondo breakdown quantum
critical point (QCP) gives rise to a divergent Gr\"uneisen ratio
with an anomalous exponent $0.7$. In particular, we fit the
experimental data of $YbRh_{2}(Si_{0.95}Ge_{0.05})_{2}$ for
specific heat, thermal expansion, and Gr\"uneisen ratio based on
our simple analytic expressions. A reasonable agreement between
the experiment and theory is found for the temperature range
between 0.4 K and 10 K. We discuss how the Gr\"uneisen ratio is a
key  measurement to discriminate between the Kondo breakdown and
spin-density wave theories.
\end{abstract}


\maketitle

Heavy-fermion quantum criticality is a typical example of a
quantum system where both strong correlations and Fermi surface
effects play a major role \cite{RMP_HFQCP}. The standard model of
quantum criticality in a metallic system is a $z = 2$ critical
theory, often referred as Hertz-Moriya-Millis theory \cite{HMM},
where $z$ is the dynamical exponent relating the variation of the
energy with the momentum, $\omega \sim q^{z}$. Unfortunately, many
heavy fermion compounds have been shown not to follow the
spin-density-wave (SDW) theoretical framework
\cite{LGW_F_QPT_Nature,INS_Local_AF,dHvA,Hall}.

An interesting suggestion is that the heavy-fermion quantum
transition is analogous to the Mott transition
\cite{DMFT,Senthil_Vojta_Sachdev,Pepin_KBQCP,Paul_KBQCP}. The
arguments in support of this view are the divergence of the
effective mass near the QCP \cite{dHvA} and the presence of
localized magnetic moments at the transition towards magnetism
\cite{INS_Local_AF}. Combined with the Fermi surface
reconstruction at the QCP \cite{dHvA,Hall}, this quantum
transition is assumed to show a breakdown of the Kondo effect as
an orbital selective Mott transition, where only the f-electrons
experience the metal-insulator transition.

Recently, this problem has been re-visited in the slave-boson
context \cite{Pepin_KBQCP,Paul_KBQCP}. The main idea is that the
Kondo breakdown QCP is multi-scale. Dynamics of hybridization
(holon) fluctuations is governed by spinon-electron polarization.
An important observation is that there should exist a Fermi
surface mismatch $q^{*} = |k_{F}^{f} - k_{F}^{c}|$ between Fermi
momentum $k_{F}^{f}$ for spinons and $k_{F}^{c}$ for conduction
electrons since fillings of spinons and electrons differ from each
other. Fermi surface mismatch gives rise to an energy gap $E^{*}$
for such spinon-electron fluctuations. Although it depends on the
value of $q^{*}$, this energy scale is shown to vary from ${\cal
O}(10^{0})$ $mK$ to ${\cal O}(10^{2})$ $mK$. When $E < E^{*}$,
holon fluctuations are undamped without considering gauge
fluctuations, thus described by $z = 2$ dilute Bose gas model. On
the other hand, when $E > E^{*}$, holon fluctuations are
dissipative since spinon-electron excitations are Landau damped,
thus described by $z = 3$ critical theory. Based on the $z = 3$
quantum criticality, recent studies \cite{Pepin_KBQCP,Paul_KBQCP}
have found quasi-linear electrical transport and logarithmically
divergent specific heat coefficient in $d = 3$, consistent with an
experiment \cite{LGW_F_QPT_Nature}.

In this paper we study the Gr\"uneisen ratio (GR)
$\Gamma_{s}(r,T) \equiv {\alpha_{s}(r,T)}/{c_{s}(r,T)}$ based on
the multi-scale Kondo breakdown QCP scenario
\cite{Pepin_KBQCP,Paul_KBQCP}, where $\alpha_{s}(r,T) =
\frac{1}{P_{0}v} \frac{\partial^{2} f_{s}(r,T)}{\partial r
\partial T}$ and $c_{s}(r,T) = - T \frac{\partial^{2} f_{s}(r,T)}{\partial T^{2}}$
are thermal expansion and molar specific heat with molar volume
$v$, respectively. Recently, GR has been proposed as one possible
measure characterizing the nature of a QCP \cite{GR_Theory}. A
remarkable feature is that the GR diverges at any QCP with an
anomalous exponent depending on the nature of the quantum
transition. Consider the scaling expression $f_{s}(r,T) =
b^{-(d+z)} f_{r}(r b^{1/\nu}, T b^{z})$ for the free energy near a
QCP at spacial dimension $d$, where $r \approx
\frac{P-P_{c}}{P_{0}}$ is a distance from the QCP ($P_{c}$) with a
pressure-unit constant $P_{0}$ and $b$ is a scaling parameter with
a correlation-length exponent $\nu$. Evaluating the thermal
expansion and molar specific heat, we find the following scaling
expressions : \[ \Gamma_{s}(r;T\rightarrow 0) = - G_{r} \ [v(P -
P_{c})]^{-1} \ , \] with $G_{r} = \nu(d - y_{0}z)/y_{0}$ where
$y_{0}$ is an exponent associated with the third law of
thermodynamics \cite{GR_Theory}, and \[\Gamma_{s}(T;r=0) = - G_{T}
\ T^{- \frac{1}{\nu z}}\ , \] with $G_{T} = \frac{1}{P_{0}v}
\frac{z[\nu(d+z)-1]}{\nu d (d+z)} \frac{[\partial
f_{r}(t,1)/\partial t]_{t=0}}{f_{r}(0,1)}$. The GR at the QCP,
$\Gamma_{s}(T;r=0)$ exhibits the scaling exponent $x =
\frac{1}{\nu z}$ in any dimension.


In a recent measurement \cite{GR_Exp} it was reported that for
$YbRh_{2}(Si_{0.95}Ge_{0.05})_{2}$, the specific heat coefficient
can be well fitted by $\gamma(T) = C(T)/T \propto \ln \Bigl(
\frac{T_{\gamma}}{T} \Bigr)$ for $\sim 0.3 K < T < 10 K$, with an
energy scale $T_{\gamma} \approx 30 K$ identified with its Kondo
temperature, and $\gamma(T) \propto T^{-1/3}$ for $T < \sim 0.3
K$. The thermal expansion coefficient was fitted as $\alpha(T)/T
\propto - \ln \Bigl( \frac{T_{\alpha}}{T} \Bigr)$ for $1 K < T <
10 K$ with a temperature scale $T_{\alpha} \approx 13 K$, and
$\alpha(T)/T \propto a_{0} + a_{1}/T$ for $\sim 0.1 K < T < 1 K$
with $a_{0} \approx 3.4 \times 10^{-6} K^{-2}$ and $a_{1} \approx
1.34 \times 10^{-6} K^{-1}$. Finally, the experiment shows that
the GR diverges with an exponent $x \sim 0.7 \pm 0.1$. This
invalidates the SDW scenario, since we have $x = 1$ owing to $z =
2$ and $\nu = 1/2$, where this critical theory is beyond its upper
critical dimension in $d \geq 2$.

\begin{figure}[t]
\vspace{5cm} \includegraphics{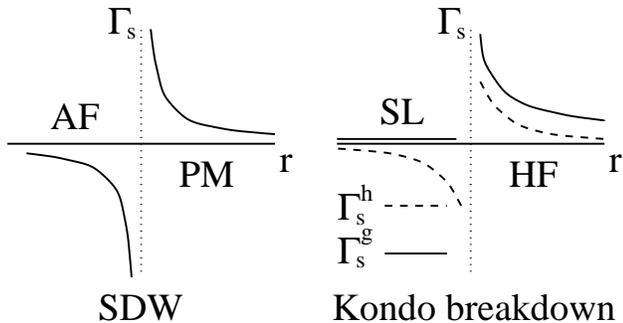} \caption{ (Color online)
Schematic diagram of Gr\"uneisen ratio $\Gamma_{s}(r;T\rightarrow
0)$ for both the SDW and Kondo breakdown scenarios, where AF and
PM represent antiferromagnetic and paramagnetic phases in the SDW
context, and SL and HF denote spin liquid and heavy fermion states
in the Kondo breakdown scenario. Two kinds of boson excitations,
hybridization ($\Gamma_{s}^{h}$) and gauge ($\Gamma_{s}^{g}$)
fluctuations, contribute to Gr\"uneisen ratio in the Kondo
breakdown scenario. Gauge fluctuations play an important role for
thermal expansion in the HF phase more than in the SL phase. As a
result, asymmetry is expected to appear for Gr\"uneisen ratio
around the Kondo breakdown QCP.} \label{fig1}
\end{figure}

In this study we show that the scenario of multi-scale Kondo
breakdown QCP gives rise to a divergent GR with the exponent
$0.7$. In particular, we fit the experimental data \cite{GR_Exp}
of $YbRh_{2}(Si_{0.95}Ge_{0.05})_{2}$ for specific heat, thermal
expansion, and GR with simple analytic expressions [Eqs. (3) and
(4)], for which the asymptotic behavior is summarized in Table I.
The $z = 3$ quantum criticality in $d = 3$ turns out to play an
essential role for thermodynamics near the QCP of
$YbRh_{2}(Si_{0.95}Ge_{0.05})_{2}$.
\begin{table}[ht]
\begin{tabular}{cccc}
 $\alpha_{s} (T)$  \; \;  & $c_{s} (T)$ \; \; & $\Gamma_{s} (T)$ \; \; \nn \hline $T^{1/3}$\; \; \;
& $- T \ln T$\; \; \;  & $-T^{- 2/3}/ \ln T $ \; \; \;  \nn \hline
\end{tabular}
\caption{Thermodynamics in the $z = 3$ regime ($d = 3$)}
\end{table}

Interestingly, varying an external parameter gives an opportunity
to distinguish the Kondo breakdown from the SDW scenario (see Fig.
1). It has been shown that the thermal expansion coefficient
should change sign across the SDW QCP in the zero temperature
limit, and GR also does accordingly \cite{GR_Theory}. In the Kondo
breakdown scenario two kinds of collective excitations,
hybridization and gauge fluctuations, contribute to thermal
expansion. Hybridization fluctuations give rise to the same sign
change as the SDW fluctuations while gauge fluctuations do not.
Considering that gauge fluctuations should remain gapless in the
spin liquid phase due to gauge invariance, their contribution for
thermal expansion is vanishingly small in the spin liquid phase.
On the other hand, they contribute to thermal expansion heavily,
approaching the QCP in the heavy-fermion phase owing to the
Anderson-Higgs mechanism. Taking into account both hybridization
and gauge fluctuations, an asymmetric feature of GR is expected to
appear around the Kondo breakdown QCP.

We start from the U(1) slave-boson representation of the Anderson
lattice model in the large-$U$ limit \bqa && L =  \sum_{i}
c_{i\sigma}^{\dagger}(\partial_{\tau} - \mu)c_{i\sigma} - t
\sum_{\langle ij \rangle} (c_{i\sigma}^{\dagger}c_{j\sigma} +
H.c.) \nn && + V \sum_{i} (b_{i}f_{i\sigma}^{\dagger}c_{i\sigma} +
H.c.) + \sum_{i}b_{i}^{\dagger} \partial_{\tau} b_{i} \nn && +
\sum_{i}f_{i\sigma}^{\dagger}(\partial_{\tau} +
\epsilon_{f})f_{i\sigma} + \frac{J}{N} \sum_{\langle ij \rangle} (
f_{i\sigma}^{\dagger}\chi_{ij}f_{j\sigma} + H.c.) \nn && + i
\sum_{i} \lambda_{i} (b_{i}^{\dagger}b_{i} + f_{i\sigma}^{\dagger}
f_{i\sigma} - 1) + \frac{J}{N} \sum_{\langle ij \rangle}
|\chi_{ij}|^{2} . \eqa Here, $c_{i\sigma}$ and $d_{i\sigma} =
b_{i}^{\dagger}f_{i\sigma}$ are conduction electron with a
chemical potential $\mu$ and localized electron with an energy
level $\epsilon_{f}$ respectively, where $b_{i}$ and $f_{i\sigma}$ are holon
and spinon, associated with hybridization and spin fluctuations.
The spin-exchange term for the localized orbital is
introduced for competition with the hybridization term, and
decomposed via exchange hopping processes of spinons, where
$\chi_{ij}$ is a hopping parameter for the decomposition.
$\lambda_{i}$ is a Lagrange multiplier field to impose the single
occupancy constraint $b_{i}^{\dagger}b_{i} + f_{i\sigma}^{\dagger}
f_{i\sigma} = N/2$, where $N$ is the number of fermion flavors
with $\sigma = 1, ..., N$.

The slave-boson mean-field analysis has shown an orbital selective
Mott transition \cite{Nature_SL} as breakdown of Kondo effect at
$J \approx T_{K}$, where $T_{K} = D
\exp\Bigl(\frac{\epsilon_{f}}{N \rho_{c}V^{2}}\Bigr)$ is the Kondo
temperature with the density of states $\rho_{c}$ for conduction
electrons \cite{Pepin_KBQCP}. If we try to understand the GR in
this level of approximation, we find $x = 1$ for the GR exponent.
Actually, one can check that the mean-field free energy satisfies
the following scaling behavior $f_{MF}(B,T) = T^{\frac{d+z}{z}}
\mathcal{F} (BT^{-\frac{1}{\nu z}})$, where $\mathcal{F} (x)$ is
an analytic function and $B = V \langle b \rangle$ is the
effective hybridization.

Fluctuation-corrections are important at the Kondo breakdown QCP,
where both hybridization and gauge fluctuations should be taken
into account carefully. Such fluctuations are treated on an
equal footing in the Eliashberg framework, where momentum
dependence in self-energies and vertex corrections are
neglected, justified by the Migdal theorem and large $N$
approximation \cite{Pepin_KBQCP,Paul_KBQCP}.

For a systematic study of thermodynamics, we construct a
Luttinger-Ward (LW) functional in the Eliashberg framework,
composed of contributions from conduction electrons, spinons,
holons, gauge fluctuations and their self-energy parts. One can
derive self-consistent Eliashberg equations for the
self-energies from variation of the LW functional with respect to
each self-energy. Using these equations, one is allowed to
simplify the LW functional as $F_{LW} = F_{FL}^{c} + F_{FL}^{f} +
F_{b} + F_{a}$, where the first two parts represent Fermi liquid
contributions for conduction electrons and spinons while the
latter two parts express hybridization and gauge contributions,
respectively. Such fermion contributions are sub-dominant compared
with boson contributions, and they can be ignored in the low
energy limit. Accordingly, thermal expansion and specific heat can
be approximated as follows near the Kondo breakdown QCP,
$\alpha_{s}(T) \approx \alpha_{b}(T) + \alpha_{a}(T)$ and
$c_{s}(T) \approx c_{b}(T) + c_{a}(T)$, respectively. As a result,
the GR is found to be $\Gamma_{s}(T) \approx \frac{\alpha_{b}(T) +
\alpha_{a}(T)}{c_{b}(T) + c_{a}(T)}$.

The bosonic part of the free energy is given by \bqa F_{s}(T)& = &
T \sum_{i\Omega} \int \frac{d^{3}q}{(2\pi)^{3}} \ln \Bigl(q^{2} +
\gamma_{b}\frac{|\Omega|}{q} + 2 \Delta_{b}\Bigr) \nn & + & T
\sum_{i\Omega} \int \frac{d^{3}q}{(2\pi)^{3}} \ln \Bigl(q^{2} +
\gamma_{a}\frac{|\Omega|}{q} + \Delta_{a}\Bigr) + F_{c} . \eqa
Here, $\gamma_{b} = \frac{2\pi}{\alpha v_{F}^{c}}$ and $\gamma_{a}
= \frac{\pi}{\alpha v_{F}^{c}} + \frac{3\pi V^{2} \rho_{c}
f_{d}}{\alpha}$ are Landau damping coefficients for holon and
gauge fluctuations, respectively, where $v_{F}^{c}$ is the Fermi
velocity for conduction electrons, $\alpha = \frac{J\chi}{t}$ is
an effective ratio between bandwidth of each fermion sector, and
$f_{d}$ is associated with an ultra-violet cutoff for gauge
fluctuations. $\Delta_{b}$ is the mass for the hybridization
fluctuations, identifying the Kondo breakdown QCP with $\Delta_{b}
= 0$. $\Delta_{a}$ is the mass for gauge fluctuations, resulting
from Anderson-Higgs mechanism, thus related with the mass of holon
as $\Delta_{a} = \frac{3NV^{4}\rho_{c}^{3}}{4\alpha u_{b}}
\Delta_{b}$, where $u_{b}$ is the strength of local interactions
for holons, phenomenologically introduced. $F_{c}$ is the
condensation part.

Performing the frequency summation and momentum integral, we find
the specific heat and thermal expansion coefficients at the QCP,
\bqa && \frac{c_{s}(T>E^{*})}{T} = \mathcal{C}_{c} \Bigl\{
\gamma_{b} \ln\Bigl( \frac{\Lambda}{T} \Bigr) + \gamma_{a}
\ln\Bigl( \frac{\gamma_{b}\Lambda}{\gamma_{a}T} \Bigr) \Bigr\} ,
\nn && \frac{c_{s}(T<E^{*})}{T} = \mathcal{C}_{c} \Bigl\{
\gamma_{b} \ln\Bigl( \frac{\Lambda}{E^{*}} \Bigr) + \gamma_{a}
\ln\Bigl( \frac{\gamma_{b}\Lambda}{\gamma_{a}T} \Bigr) \Bigr\}
\eqa and \bqa && \frac{\alpha_{s}(T>E^{*})}{T} =
\mathcal{C}_{\alpha} \frac{\partial \Delta_{b}}{\partial P} \Bigl(
2 \gamma_{b}^{\frac{1}{3}} + \frac{3NV^{4}\rho_{c}^{3}}{4\alpha
u_{b}} \gamma_{a}^{\frac{1}{3}} \Bigr) T^{- \frac{2}{3}} , \nn &&
\frac{\alpha_{s}(T<E^{*})}{T} = \mathcal{C}_{\alpha}
\frac{\partial \Delta_{b}}{\partial P} \Bigl( 2
\gamma_{b}^{\frac{1}{3}} E^{*-\frac{2}{3}} +
\frac{3NV^{4}\rho_{c}^{3}}{4\alpha u_{b}} \gamma_{a}^{\frac{1}{3}}
T^{- \frac{2}{3}} \Bigr) , \nn \eqa where $\mathcal{C}_{c} =
\frac{4}{3\pi^{3}} \int_{0}^{\infty} {d y} \Bigl( -
\frac{y^{2}}{\sinh^{2} y } + \frac{y^{3} \coth y }{\sinh^{2} y }
\Bigr)$ and $\mathcal{C}_{\alpha} =
\frac{2^{\frac{1}{3}}}{\pi^{3}} \Bigl(\int_{0}^{\infty} d x
\frac{x^{3}}{x^{6} + 1}\Bigr) \Bigl( \int_{0}^{\infty} d y
\frac{y^{\frac{4}{3}}}{\sinh^{2} y} \Bigr)$ are positive numerical
constants. The condensation part is assumed to be almost constant
for temperature dependence, thus can be ignored for thermal
expansion. Note that there is an unknown constant $\frac{\partial
\Delta_{b}}{\partial P}$ with pressure $P$ in the thermal
expansion coefficient, determining its overall sign. Recalling
that it is negative for $YbRh_{2}(Si_{0.95}Ge_{0.05})_{2}$
\cite{GR_Exp}, we see $\frac{\partial \Delta_{b}}{\partial P} <
0$. This implies that pressure puts the QCP of
$YbRh_{2}(Si_{0.95}Ge_{0.05})_{2}$ toward the heavy-fermion side
if it is identified with the Kondo breakdown QCP. One can check
that holon thermodynamics is consistent with $z = 3$ scaling for
$T > E^{*}$ while gauge thermodynamics is for all temperatures. In
addition, both specific heat and thermal expansion coefficients
are constant for holon fluctuations at $T < E^{*}$, consistent
with Fermi liquid physics.

\begin{figure}[t]
\vspace{5cm} \includegraphics{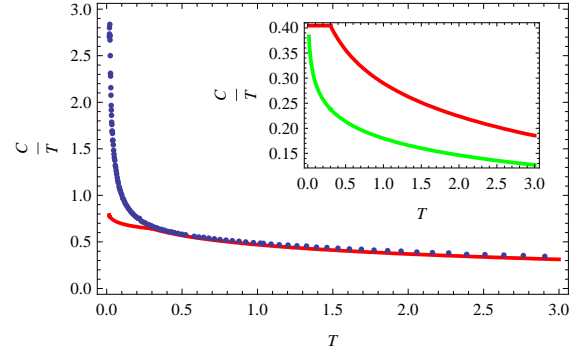} \caption{ (Color online) Specific
heat coefficient, where the blue dotted line represents an
experimental data and the red thick line does our theory. Inset:
Specific heat contributions from hybridization fluctuations (red)
and gauge fluctuations (green). Note that the specific heat
coefficient from hybridization fluctuations is twice bigger than
that from gauge fluctuations. } \label{fig2}
\end{figure}
\begin{figure}[t]
\vspace{5cm} \includegraphics{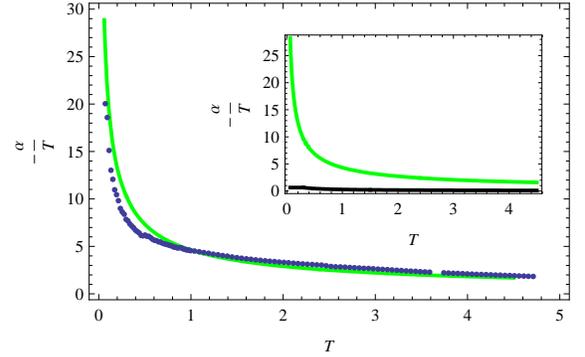} \caption{ (Color online) Thermal
expansion coefficient, where the blue dotted line represents an
experimental data and the green thick line does our theory. Inset:
Thermal expansion contributions from hybridization fluctuations
(black) and gauge fluctuations (green). Note that the thermal
expansion coefficient from gauge fluctuations is much larger than
that from hybridization fluctuations although this physics depends
on the local-interaction strength $u_{b}$ for hybridization
fluctuations. } \label{fig3}
\end{figure}
\begin{figure}[t]
\vspace{5cm} \includegraphics{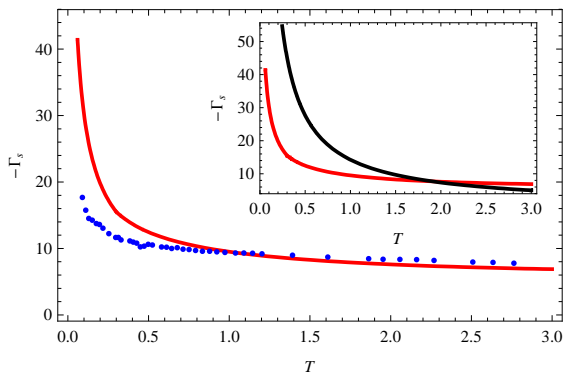} \caption{ (Color online) Gr\"uneisen
ratio, where the blue dotted line represents an experimental data
and the red thick line does our theory. Inset: Comparison between
the $d = 3$ Kondo breakdown (red) and $d = 2$ SDW (black)
theories, where the SDW theory exhibits more singular behavior at
low temperatures.} \label{fig4}
\end{figure}

Using Eqs. (3) and (4), we try to fit the experimental data of
Ref. \cite{GR_Exp}. The density of states $\rho_{c} =
\frac{1}{2D}$ for conduction electron with the bandwidth $D = 6 t
= 10^{4} K$ and the ratio between bandwidths $\alpha = 10^{-3}$
are fixed. Considering $v_{F}^{c} = \frac{k_{F}^{c}}{m_{c}}
\approx 2 t$ for the sphere Fermi surface, $\gamma_{b}$ is
determined as $\gamma_{b} = 2.693$. $V$ can be deduced from the
Kondo temperature $T_{K} \approx 30 K$ with $\epsilon_{f} = -
\frac{D}{2}$, thus $V = 0.293 D$. $f_{d}$ is used as a fitting
parameter, determining $\gamma_{a} = 1.367$. The cutoff $\Lambda$
in specific heat is set to be an effective bandwidth for localized
spins, i.e., $\Lambda \approx \alpha D$. $E^{*}$ is approximately
given by the upturn temperature for specific heat, here $E^{*}
\approx 0.3 K$. For thermal expansion, we have two free
parameters, $u_{b}$ and $\frac{\partial \Delta_{b}}{\partial P}$.

Fig. 2 shows the fitting for the specific heat coefficient
$c_{s}(T)/T$. For $T > E^{*}$, we have a very good matching unlike
for $T < E^{*}$. Although the origin of this upturn behavior is
not explained yet clearly, two dimensional ferromagnetic
fluctuations \cite{FM_Exp} may give one possible explanation,
resulting in $c_{FM}^{d=2}(T)/T \propto T^{-1/3}$. From the inset
figure, we can conclude that both hybridization and gauge
fluctuations are important for specific heat near the Kondo
breakdown QCP. Considering that the hybridization fluctuations
arise from collective excitations of conduction electrons and
spinons, and gauge fluctuations result from spinon current-current
correlations, one can expect that both fluctuations will
contribute in a similar fashion.

Fig. 3 shows the fitting for the thermal expansion coefficient
$\alpha_{s}(T)/T$, where we have a rather good agreement between
experiment and theory. Although we have used two free parameters
$u_{b}$ and $\frac{\partial \Delta_{b}}{\partial P}$, such
parameters can change only the overall scale, thus one may regard
that only one parameter is used. The inset figure exhibits that
contributions from gauge fluctuations are much larger than those
from hybridization ones. Although this physics depends on the
local-interaction strength $u_{b}$ for hybridization fluctuations,
it is valid as far as $u_{b} \ll \frac{1}{10\alpha}
\frac{V^{4}}{D^{3}}$ is satisfied in Eq. (4), i.e., in the weak
coupling limit preserving the present picture of the Kondo
breakdown QCP.
Our fitting for the
thermal expansion coefficient may be the first explicit
demonstration, supporting importance of gauge fluctuations.

Fig. 4 shows a reasonable match between experiment and theory for
the GR above the upturn temperature. As shown in the inset figure,
the $d = 2$ SDW theory shows more singular behavior at low
temperatures, deviating from the experiment more seriously.

In conclusion, we have fitted the experimental data of
$YbRh_{2}(Si_{0.95}Ge_{0.05})_{2}$ for specific heat, thermal
expansion and Gr\"uneisen ratio based on  simple analytic formulae
in the multi-scale Kondo breakdown scenario. Both hybridization
and gauge fluctuations contribute to specific heat in a similar
fashion around the QCP. Gauge fluctuations are more important in
the heavy-fermion phase than in the spin liquid phase for thermal
expansion, causing an asymmetry for Gr\"uneisen ratio around the
QCP. This feature can be used to discriminating the Kondo
breakdown scenario from the SDW framework. These $z = 3$ critical
fluctuations explain the divergent Gr\"uneisen ratio with the
anomalous exponent $0.7$ beyond the SDW theory.
We suggest that
two dimensional ferromagnetic fluctuations may give one possible
explanation for thermodynamics in the low temperature region below
$E^{*}$, not captured in the present framework.

We thank F. Steglich and N. Oeschler for sending us the original
data of their experiment. K.-S. Kim thanks I. Paul and A. Cano for
helpful discussions. This work is supported by the French National
Grant ANR36ECCEZZZ. K.-S. Kim is also supported by the Korea
Research Foundation Grant (KRF-2007-357-C00021) funded by the
Korean Government.


\begin{thebibliography}{9}
\bibitem{RMP_HFQCP} P. Gegenwart, Q. Si, and F. Steglich,
Nature Physics {\bf 4}, 186 (2008); H. v. Lohneysen, A. Rosch, M.
Vojta, and P. Wolfle, Rev. Mod. Phys. {\bf 79}, 1015 (2007).
\bibitem{HMM} T. Moriya and J. Kawabata, J. Phys. Soc. Jpn. {\bf
34}, 639 (1973); T. Moriya and J. Kawabata, J. Phys. Soc. Jpn.
{\bf 35}, 669 (1973); J. A. Hertz, Phys. Rev. B {\bf 14}, 1165
(1976); A. J. Millis, Phys. Rev. B {\bf 48}, 7183 (1993).
\bibitem{LGW_F_QPT_Nature} J. Custers, P. Gegenwart, H. Wilhelm,
K. Neumaier, Y. Tokiwa, O. Trovarelli, C. Geibel, F. Steglich, C.
Pepin, and P. Coleman, Nature {\bf 424}, 524 (2003).
\bibitem{INS_Local_AF} A. Schroder, G. Aeppli, R. Coldea, M. Adams,
O. Stockert, H.v. Lohneysen, E. Bucher, R. Ramazashvili, and P.
Coleman, Nature {\bf 407}, 351 (2000).
\bibitem{dHvA} H. Shishido, R. Settai, H. Harima, and Y. Onuki,
J. Phys. Soc. Jpn. {\bf 74}, 1103 (2005).
\bibitem{Hall} S. Paschen, T. Luhmann, S. Wirth, P. Gegenwart,
O. Trovarelli, C. Geibel, F. Steglich, P. Coleman, and Q. Si,
Nature {\bf 432}, 881 (2004).
\bibitem{DMFT} L. De Leo, M. Civelli, and G. Kotliar,
arXiv:0804.3314 (unpublished).
\bibitem{Senthil_Vojta_Sachdev} T. Senthil, M. Vojta, and
S. Sachdev, Phys. Rev. B {\bf 69}, 035111 (2004).
\bibitem{Pepin_KBQCP} C. Pepin, Phys. Rev. Lett. {\bf 98}, 206401
(2007); Phys. Rev. B {\bf 77}, 245129 (2008).
\bibitem{Paul_KBQCP} I. Paul, C. Pepin, and M. R. Norman,
Phys. Rev. Lett. {\bf 98}, 026402 (2007); I. Paul, C. Pepin, M. R.
Norman, Phys. Rev. B {\bf 78}, 035109 (2008).
\bibitem{GR_Theory} L. Zhu, M. Garst, A. Rosch, and Q. Si,
Phys. Rev. Lett. {\bf 91}, 066404 (2003).
\bibitem{GR_Exp} R. Kuchler, N. Oeschler, P. Gegenwart,
T. Cichorek, K. Neumaier, O. Tegus, C. Geibel, J. A. Mydosh, F.
Steglich, L. Zhu, and Q. Si, Phys. Rev. Lett. {\bf 91}, 066405
(2003).
\bibitem{Nature_SL} The Mott insulating phase in the present
scenario is U(1) spin liquid with a spinon Fermi surface. Its
stability can be argued in the large flavor approximation,
consistent with the Eliashberg framework. Experimentally, rather
large entropy and small magnetic moments in the antiferromagnetic
phase seem to support the presence of gauge fluctuations and the
fact that such antiferromagnetism may arise from the spin liquid
state.
\bibitem{FM_Exp} K. Ishida, K. Okamoto, Y. Kawasaki, Y. Kitaoka,
O. Trovarelli, C. Geibel, and F. Steglich, Phys. Rev. Lett. {\bf
89}, 107202 (2002).
\end{thebibliography}
\end{document}